\documentclass[%
 aps,
 pre,%
 amsmath,amssymb,floatfix,superscriptaddress,
 reprint,%
nofootinbib
]{revtex4-2}
\usepackage{nameref}
\usepackage{soul}
\usepackage{hyperref}

\usepackage{graphicx}
\usepackage{bm}
\usepackage{color}
\usepackage{multirow}

\definecolor{mauricio}{RGB}{21,168,48}

\begin{document}

\title{
Subsampled directed-percolation models
explain scaling relations experimentally observed in the brain}

\author{Tawan T. A. Carvalho}
\author{Antonio J. Fontenele}
\affiliation{Departamento de F\'isica, Universidade Federal de Pernambuco, Recife, PE, 50670-901, Brazil}
\author{Mauricio Girardi-Schappo}
\affiliation{Physics Department, STEM Complex, University of Ottawa, Ottawa, ON, K1N 6N5, Canada}
\affiliation{Departamento de F\'isica, FFCLRP, Universidade de S\~ao Paulo, Ribeir\~ao Preto, SP, 14040-901, Brazil}
\author{Tha\'is Feliciano}
\affiliation{Departamento de F\'isica, Universidade Federal de Pernambuco, Recife, PE, 50670-901, Brazil}
\author{Leandro A. A. Aguiar}
\affiliation{Departamento de Ci\^encias Fundamentais e Sociais, Universidade Federal da Paraíba, Areia, PB, 58397-000, Brazil}
\author{Thais P. L. Silva}
\affiliation{Departamento de F\'isica, Universidade Federal de Pernambuco, Recife, PE, 50670-901, Brazil}
\author{Nivaldo A. P. de Vasconcelos}
\affiliation{Departamento de F\'isica, Universidade Federal de Pernambuco, Recife, PE, 50670-901, Brazil}
\affiliation{Life and Health Sciences Research Institute (ICVS), School of Medicine, University of Minho, Braga, 4710-057, Portugal}
\affiliation{ICVS/3B’s - PT Government Associate Laboratory, Braga/Guimar\~aes, Portugal}
\author{Pedro V. Carelli}
\affiliation{Departamento de F\'isica, Universidade Federal de Pernambuco, Recife, PE, 50670-901, Brazil}
\author{Mauro Copelli}
\email{mauro.copelli@ufpe.br}
\affiliation{Departamento de F\'isica, Universidade Federal de Pernambuco, Recife, PE, 50670-901, Brazil}

\date{\today}
\begin{abstract}
Recent experimental results on spike avalanches measured in the urethane-anesthetized rat cortex have revealed scaling relations that indicate a phase transition at a specific level of cortical firing rate variability.
The scaling relations point to critical exponents whose values differ from those of a branching process, which has been the canonical model employed to understand brain criticality.
This suggested that a different model, with a different phase transition, might be required to explain the data. 
Here we show that this is not necessarily the case. 
By employing two different models belonging to the same universality class as the branching process (mean-field directed percolation) and treating the simulation data exactly like experimental data, we reproduce most of the experimental results. 
We find that subsampling the model and adjusting the time bin used to define avalanches (as done with experimental data) are sufficient ingredients to change the apparent exponents of the
critical point.
Moreover, experimental data is only reproduced within a very narrow
range in parameter space around the
phase transition.

\end{abstract}

\maketitle

%

\newpage

\section{Introduction}

In the first results that fueled the critical brain hypothesis, Beggs and Plenz~\citep{beggs2003neuronal} observed intermittent bursts of local field potentials (LFPs) in \textit{in vitro} multielectrode recordings of cultured slices of the rat brain. 
Events occurred with a clear separation of time scales, and were named neuronal avalanches. 

A neuronal avalanche can be characterized by its size $S$, which is the total number of events recorded by electrodes between periods of silence, and by its duration $T$, which is the number of consecutive time bins spanned by an avalanche. 
Beggs and Plenz found power-law distributions for the sizes of avalanches, 
\begin{equation}
\label{eq:P(S)}
P(S)\sim S^{-\tau} \; ,
\end{equation}
with $\tau \simeq 3/2$, and suggested, based on their data,
a power-law distribution of avalanche duration,
\begin{equation}
\label{eq:P(T)}
P(T)\sim T^{-\tau_t}\; ,    
\end{equation}
with $\tau_t = 2$.
These scale-invariant distributions were interpreted as a signature that the 
brain could be operating at criticality -- a second-order phase 
transition~\citep{beggs2003neuronal,beggs2007criticality,chialvo2010emergent,shew2013functional}. 
In particular, these two critical exponents together are
compatible with a branching process at 
its critical point~\citep{Harris1963branching}, thus pointing to a phase transition between a so-called absorbing phase (zero population firing rate) and an active phase (non-zero stationary population firing rate). 

Due to its appeal, simplicity and familiarity within the statistical physics community, the critical branching process has become a canonical model for understanding criticality in the brain.
In fact, these exponents are compatible with a larger class of models, namely, any model belonging to the mean-field directed percolation (MF-DP) universality class~\citep{Munoz99}.
In the theory of critical phenomena, two models which can be different in their details are said to belong to the same universality class when the critical exponents which characterize their phase transition coincide~\citep{Binney92}. 
In general, probabilistic contagion-like models which have a unique absorbing state (all sites ``susceptible'' or, in the neuroscience context, all neurons quiescent) and no further symmetries tend to belong to the directed-percolation universality class~\citep{Janssen1981,Grassberger1982,Marro99}. 
If the network has topological dimension above 4 (such as random or complete graphs),
the model usually belongs to the MF-DP universality class. 

More recent experimental results, however, challenged the MF-DP scenario proposed by Beggs and Plenz~\citep{beggs2003neuronal}. 
For instance, avalanche exponents in ex-vivo recordings of the turtle
visual cortex deviated significantly from $\tau=3/2$ and
$\tau=2$~\citep{Shew15}. 
Discrepancies in exponent values were also observed in spike avalanches of 
rats under ketamine-xylazine anesthesia~\citep{ribeiro2010spike} and M/EEG 
avalanches in resting or behaving 
humans~\citep{PalvaPNAS13,Zhigalov2015}, among others.  

Furthermore, Touboul and Destexhe~\citep{touboul2017power} argued that the 
power-law signature alone
in the distributions of size (Equation~\ref{eq:P(S)}) and duration (Equation~\ref{eq:P(T)}) of avalanches is insufficient to claim criticality, since power laws can be observed in non-critical models as well. 
They suggested that another scaling relation should be tested as a stronger criterion. 
This was based on the result that at criticality the average avalanche size $\langle S \rangle$ for a given duration $T$ must obey
\begin{equation}
\label{eq:S-vs-T}
\langle S \rangle \sim T^{\frac{1}{\sigma \nu z}}\; ,    
\end{equation}
where $1/(\sigma \nu z)$ is a combination of critical exponents
that at criticality satisfy the so-called crackling noise scaling 
relation~\citep{Munoz99,Sethna2001crackling,friedman2012}
\begin{equation}
\label{eq:scaling}
\frac{1}{\sigma\nu z}=\frac{\tau_t-1}{\tau-1}\; .
\end{equation}

Equation~\eqref{eq:scaling} is  a stronger criterion for criticality because it is expected not to be satisfied by non-critical models~\citep{touboul2017power}. 
In the MF-DP case,
the avalanche exponents obey $(\tau_t-1)/(\tau-1)=2$ and $1/(\sigma\nu z)=2$, independently.
The absolute difference between the two sides of Equation~\eqref{eq:scaling} can even be employed as a metric for the distance to criticality~\citep{Ma2019}, or to identify criticality in more general phase transitions of
neuronal networks~\citep{GirardiV12018}.

Recently, Fontenele et al.~\citep{fontenele2019criticality} investigated cortical spike avalanches of urethane-anesthetized rats using these equations. 
This experimental setup is known to yield spiking activity which is highly variable, ranging from strongly asynchronous to very synchronous population
activity~\citep{clement2008cyclic}. The synchronicity regimes can be characterized by different ranges of the coefficient of variation ($CV$) of the instantaneous population firing rate~\citep{Vasconcelos2017}, which is used as a proxy to the cortical state, in a time resolution of seconds~\citep{harris2011cortical}.  
By parsing the data according to levels of spiking variability, Fontenele et al.~\citep{fontenele2019criticality} found that the scaling relation in Equation~\eqref{eq:scaling} was only satisfied at an intermediate value of $CV$.
They suggested that a critical point is then present in cortical activity
away from both the desynchronized and synchronized ends of the spiking
variability spectrum.
However, the critical exponents were, again, not compatible (within error bars) with MF-DP values: 
$\langle \tau \rangle \simeq 1.52 \pm 0.09$, $\langle \tau_t \rangle \simeq 1.7 \pm 0.1$ and $\langle 1/(\sigma \nu z)\rangle \simeq 1.28 \pm 0.03$. 
They interpreted those results as an incompatibility with the theoretical MF-DP scenario~\citep{fontenele2019criticality}.

Here we revisit this issue by studying the data produced by two theoretical models in the MF-DP universality class under the same conditions as those of experimental data.
Despite the large number of simulated neurons ($\sim10^5$), we intentionally restrict the
theoretical analysis to a small subset of cells ($\sim10^2$), mimicking the fact that one can
only record a few hundred neurons among the millions that comprise the rat's brain -- a technique called subsampling.
Different groups have shown that subsampling alters the apparent
distribution of avalanches~\citep{Priesemann09,ribeiro2010spike,Girardi2013,Ribeiro14a, Priesemann2014subcriticalbrain,Levina2017subsampling,Wilting2019criticality}.
We show that combining subsampling with the experimental pipeline
reconcile the empirical power-law avalanches with the theoretical MF-DP universality class.

\section{Methods} \label{Sec:method}
\subsection{A spiking neuronal network with excitation and inhibition}
\label{sec:GGL}

We use the excitatory/inhibitory network of Girardi-Schappo et al.~\citep{Girardi2020bal}, where each neuron is a stochastic leaky integrate-and-fire unit with discrete time step equal to 1~ms, connected
in an all-to-all graph. 
A binary variable indicates if the neuron fired ($X(t)=1$) or not ($X(t)=0$). 
The membrane potential of each cell $i$ in either the excitatory ($E$) or inhibitory ($I$) population is given by
\begin{multline}
\label{GGLModel}
V_i^{E/I}(t+1) = \Bigg[ \mu V_i^{E/I}(t) + I_e
+\dfrac{J}{N} \sum_{j=1}^{N_E} X_j^E(t)\\
-\dfrac{gJ}{N} \sum_{j=1}^{N_I} X_j^I(t) \Bigg]
\bigg(1-X_i^{E/I}(t) \bigg)\; ,
\end{multline}
where $J$ is the synaptic coupling strength, $g$ is the inhibition to excitation (E/I)
coupling strength ratio, $\mu$ is the leak time constant, and $I_e$ is an external current.
The total number of neurons in the network is $N=N_E+N_I=10^5$, where the fractions of excitatory and inhibitory neurons are kept fixed at $p=N_E/N=0.8$ and $q=N_I/N=0.2$, respectively, as reported for cortical data~\citep{Somogyi1998}.
Note that the membrane potential is reset to zero in the time step 
following a spike.


At any time step, a neuron fires according to 
a piecewise linear sigmoidal probability $\Phi(V)$,
\begin{multline}
\Phi(V) \equiv P\left(X=1|V\right)
\\= \Gamma\:(V-\theta)\:\Theta(V-\theta)\:
\Theta(V_S - V) + \Theta(V-V_S),
\label{phifunc}
\end{multline}
where $\theta=1$ is the firing threshold, $\Gamma$ is the firing gain
constant, $V_S=1/\Gamma+\theta$ is the saturation potential, and
$\Theta(x>0)=1$ (zero
otherwise) is the step function. 
For simplicity, the parameter $\mu=0$ is chosen without lack
of generality, since it
does not change the phase transition of the model~\citep{Girardi2020bal}.
The external current $I_e>V_S$ is used only to spark a new avalanche in a single
excitatory neuron when the network activity dies off (it is kept as
$I_e=\theta$ otherwise).

This model is known to present a directed percolation critical point~\citep{Girardi2020bal} at $g_c=p/q-1/(q\Gamma J)=1.5$ (for $\Gamma=0.2$ and $J=10$), such that $g<g_c$ is the active excitation-dominated (supercritical) phase and $g>g_c$ corresponds to the  inhibition-dominated absorbing state (subcritical). 
The synapses in the critical point $g_c$ are dynamically balanced: fluctuations in excitation are immediately followed by counter fluctuations in inhibition~\citep{Girardi2020bal}.
The initial condition of the simulations has all neurons quiescent except
for a seed neuron to spark activity.
This procedure was repeated whenever the system went back to the absorbing state. 

\subsection{Probabilistic cellular automaton model}
\label{sec:KC}

The robustness of our findings was cross-checked using probabilistic cellular automata in a random network~\citep{kinouchi2006optimal}. 
This model closely resembles a standard branching process and is known to mimic the changing inhibition-excitation levels of cortical cultures~\citep{shew2009neuronal}.

Each site $i$ ($i = 1, ..., N$) has 5 states: the silent state, $s_i = 0$, the active state, $s_i = 1$, corresponding to a spike, and the remaining three
states, $s_i=2,3,4$, in which the site will not respond to incoming stimuli (refractory states). 
Each site receives input from
$K$ presynaptic neighbors which are randomly selected at the start and kept fixed throughout the simulations.
A quiescent site $i$ becomes excited ($s_i(t) = 0 \to s_i(t+1) = 1$) with probability $p_{ij}$ if a presynaptic neighbor $j$ is active at time $t$. 
All presynaptic neighbors are swept and independently considered at each time step, so that 
\begin{multline}
    P\left(s_i(t+1)=1|s_i(t)=0\right) = \\
    1 - (1-h_i)
    \prod_{j\in {\cal N}(i)}^K
    \left[ 1-p_{ij}s_j(t) \right]
    \; ,
\end{multline}
where $h_i$ is the probability of unit $i$ spiking due to an external stimulus and $\mathcal{N}(i)$ is the set of presynaptic neighbors of $i$.
The remaining transitions happen with probability 1, including
the transition $4\to0$ that returns the site to its initial quiescent state.
The time step of the model corresponds to 1~ms.

We initially choose the  random variables $\{p_{ij}\}$ from a uniform distribution in the interval
[0, $2 \lambda / K$].
The so-called branching ratio $\lambda = K\langle p_{ij} \rangle$
is the control parameter of the model. 
This model undergoes a MF-DP phase transition at $\lambda=\lambda_c = 1$~\citep{kinouchi2006optimal}.
For $\lambda < 1$, the system is in the subcritical phase and eventually reaches the absorbing state ($s_i=0, \forall i$). 
For $\lambda > 1$, the system presents self-sustained activity, i.e. a nonzero stationary density of population firings (the supercritical phase).

In our simulations we used $K=10$ neighbors for each of the $N = 10^5$ sites.
Similarly to the spiking neuronal network model, a single random neuron was stimulated
($h_i=1$) only when the system reached the absorbing state, sparking the network activity
and subsequently being set back to $h_i=0$.
The initial condition was set with a single randomly chosen site active and the others in the silent state.

\subsection{Experimental data acquisition}
\label{sec:dat}

We used 5 rats Long-Evans (\textit{Rattus norvegicus}) (male, 280-360~g, 2-4 months old). They were obtained from the animal house of the Laboratory of Computational and Systems Neuroscience, Department of Physics, Federal University of Pernambuco (UFPE).

The rats were maintained in the light/dark cycle of 12 h and their water and food were \textit{ad libitum}. The experimental protocol was approved by the Ethics Committee on Animal Use (CEUA) of UFPE (CEUA: 23076.030111/2013-95 and 12/2015), in accordance with the basic principles for research animals established by the National Council for the Control of Animal Experimentation (CONCEA).

The animals were anesthetized with urethane (1.55~g/kg), diluted at 20\% in saline, in 3 intraperitoneal (i.p.) injections, 15~min apart. The rats were placed in a stereotaxic frame and the coordinates to access the primary visual cortex (V1) were marked (Bregma: AP = -7.2, ML = 3.5) \citep{paxinos2007rat}.  A cranial window in the scalp was opened at this site with
an area of approximately 3~mm\textsuperscript{2}.

In order to record extra-cellular voltage, we used a 64-channels  multielectrode silicon probe (Neuronexus technologies, Buzsaki64spL-A64). This probe has 60 electrodes disposed in 6 shanks separated by 200~$\mu$m, 10 electrodes per shank with impedance of 1--3~MOhm at 1~kHz. Each electrode has 160~$\mu$m$^{2}$ and they are in staggered positions  20~$\mu$m apart. 

The acquired data was sampled at 30~KHz, amplified and digitized in a single head-stage (Intan RHD2164)~\citep{siegle2017open}. 
We recorded spontaneous activity, during long periods ($\geq$ 3 hours). We used the open-source software Klusta to perform the automatic spike sorting on raw electrophysiological data \citep{rossant2016spike}. 
The automatic part is divided in two major steps, spike detection and automatic clustering. The first step detects action potentials and the second one arrange those spikes into clusters according to their similarities (waveforms, PCA, refractory period).
After the automatic part, all formed clusters are reanalyzed using the graphic interface phy kwikGUI \footnote{https://github.com/cortex-lab/phy}. 
Manual spike sorting allows the identification of each cluster of neuronal activity as single-unit activity (SUA) or multi-unit activity (MUA). 
We use both SUA and MUA clusters for our study.

\subsection{Avalanche analysis with CV parsing}
\label{sec:cvparsing}

To study neuronal avalanches at different levels of spiking variability, we segment both the
neurophysiological and simulated data in non-overlapping windows of width $w=10$~s
(unless otherwise stated). 
Each of these 10~s epochs is subdivided in non-overlapping intervals $\{\zeta_j\}$
of duration $\Delta T=50$~ms (unless otherwise stated) in which we estimate the population
spike-count rate $R_j$.
We then calculate the coefficient of variation ($CV$) for the $i$-th 10~s window: 
\begin{equation}
\label{CV}
CV_i = \frac{\sigma_i}{\mu_i}\,,
\end{equation}
where $CV$ is dimensionless, and $\sigma_{i}$ and $\mu_{i}$ correspond
to the standard deviation and the mean of $\{R_j\}$, respectively.

For each 10~s window with a particular $CV$ level,
we proceed with the standard avalanche analysis of Beggs and Plenz~\citep{beggs2003neuronal}. 
The summed population activity is sliced in non-overlapping temporal bins
of width $\Delta t=\langle ISI \rangle$ (the average inter-spike interval).
Population spikes preceded and followed by silence define a spike avalanche.  
The number of spikes correspond to the avalanche size $S$, whereas the number of time bins
spanned by the avalanche is its duration $T$.
Following this methodology, we associate each 10~s $CV_i$ window with
its corresponding set of $n_i$ avalanche sizes
$\mathbf{S_i}\equiv \left \{ S_{i1}, S_{i2}, ...,S_{in_i} \right \}$ 
and durations
$\mathbf{T_i}\equiv \left \{ T_{i1}, T_{i2}, ...,T_{in_i} \right \}$.

To estimate the avalanche exponents $\tau$ and $\tau_t$, we first ranked the sets $\{\mathbf{S_i}\}$ and $\{\mathbf{T_i}\}$ according to their $CV$ values. 
Next, in order to increase the number of samples while preserving the level of spiking variability, we pooled $NB$ consecutive ranked blocks of similar $CV$ values ($NB = 50$ unless otherwise stated). 
For each set of $NB$ blocks we calculated the average coefficient of variation $\langle CV \rangle$. %
The exponents of the size and duration distributions were obtained  via  a  Maximum  Likelihood  Estimator (MLE)  procedure~\citep{Klaus14,Corral2013,Marshall2016analysis}
on a discrete power-law distribution  
\begin{equation}
\label{eq:pmf}
f(x) = \frac{1}{\sum_{x=x_{min}}^{x_{max}}(\frac{1}{x})^\alpha}\left(\frac{1}{x}\right)^\alpha \; .
\end{equation}
The standard choice of fitting parameters, for both experimental and subsampled simulated data, was $S_{min} = 2$ and $S_{max} = 100$ for size distributions and $T_{min} = 2$ and $T_{max} = 30$ for duration distributions. 
The exceptions to this choice were for the data shown in Figure~\ref{fig:subsampling}C and~\ref{fig:subsampling}D, due to a change of orders of magnitude in the number of neurons sampled. 
The specific parameters for these cases are shown in Table~\ref{tab:limits}.

\begin{table}[b]
\begin{center}
\caption{Limits chosen for the calculation of the $\alpha$ exponent~(Equation~\ref{eq:pmf}) via Maximum Likelihood Estimator (MLE) only for the data shown in Figure~\ref{fig:subsampling}C and D ($\Delta t = 1$~ms). See text for details.}
\label{tab:limits}
\begin{tabular}{ccccc}
\textbf{n} & \multicolumn{2}{c}{\textbf{Size distribution}} & \multicolumn{2}{c}{\textbf{Duration distribution}} \\ \cline{2-5} 
                            & \textbf{$S_{min}$}     & \textbf{$S_{max}$}    & \textbf{$T_{min}$}       & \textbf{$T_{max}$}      \\ \hline
100                         & 2                      & 30                    & 2                        & 15                      \\
200                         & 2                      & 100                   & 2                        & 50                      \\
500                         & 2                      & 200                   & 2                        & 70                      \\
1000                        & 2                      & 200                   & 2                        & 70                      \\
2000                        & 2                      & 300                   & 3                        & 100                     \\
5000                        & 2                      & 500                   & 4                        & 100                     \\
10000                       & 5                      & 3000                  & 5                        & 150                     \\
20000                       & 5                      & 5000                  & 5                        & 200                     \\
30000                       & 10                     & 10000                 & 10                       & 200                     \\
40000                       & 10                     & 10000                 & 10                       & 250                     \\
50000                       & 10                     & 10000                 & 10                       & 300                     \\
100000                      & 10                     & 20000                 & 10                       & 300                     \\ \hline
\end{tabular}
\end{center}
\end{table}

After the MLE fit we use the Akaike Information Criterion ($AIC$) as a measure of the relative quality of a given statistical model for a data set: 
\begin{eqnarray}
\label{eq:AIC} AIC & = & 2k-2\ln(\hat{L})+\frac{2k^2+2k}{N-k-1}\; ,
\end{eqnarray}
where $\hat{L}$ is the likelihood at its maximum, $k$ is number of parameters and $N$ the sample size~\citep{Akaike75}. Starting from the principle that lower $AIC$ indicates a more parsimonious model, 
we define $\Delta  \equiv  AIC_{ln}-AIC_{pl}$, where $AIC_{ln}$ and $AIC_{pl}$ correspond to the $AIC$ of a log-normal and a power-law model, respectively. 
Therefore, $\Delta> 0$ implies that a power-law model is a better fit to the data than a log-normal.
Our scaling relation analyses were restricted to distributions that satisfied $\Delta>0$. 

\subsection{Pairwise correlations}

Pairwise spiking correlations were estimated using only the SUA or
the simulated data in the following way: first, for each cell $k$ we obtain a spike count time
series $R^{(k)}(t)$ at millisecond resolution (${\Delta T = 1}$~ms),
then each spike count time series $R^{(k)}$ is convolved with a kernel $h_{t_{1},t_{2}}(t)$
to estimate the $k$-th mean firing rate $n^{(k)}(t)$:
\begin{equation}
\label{hat1}
n^{(k)}(t) = h_{t_{1},t_{2}}(t)\ast R^{(k)}(t)  \; ,
\end{equation}
where $h_{t_{1},t_{2}}(t)$ is a Mexican-hat kernel obtained by the difference between
zero-mean Gaussians with standard deviations $t_1=100$~ms
and $t_2=400$~ms~\citep{renart2010}. 
The $n_k(t)$ are used to calculate the spiking correlation coefficient
between two units $k$ and $l$:
\begin{equation}
\label{hat2}
r^{(k,l)} = \frac{\mathrm{Cov}\left ( n^{(k)},n^{(l)}\right )}{\sqrt{\mathrm{Var}\left(n^{(k)}\right)\mathrm{Var}\left(n^{(l)}\right )}}
  \;,
\end{equation}
where Var and Cov are the variance and covariance over $t$, respectively.

\section{Results}

\subsection{Avalanches in the fully sampled model}

\begin{figure*}[!tbhp]
\includegraphics[width=\linewidth]{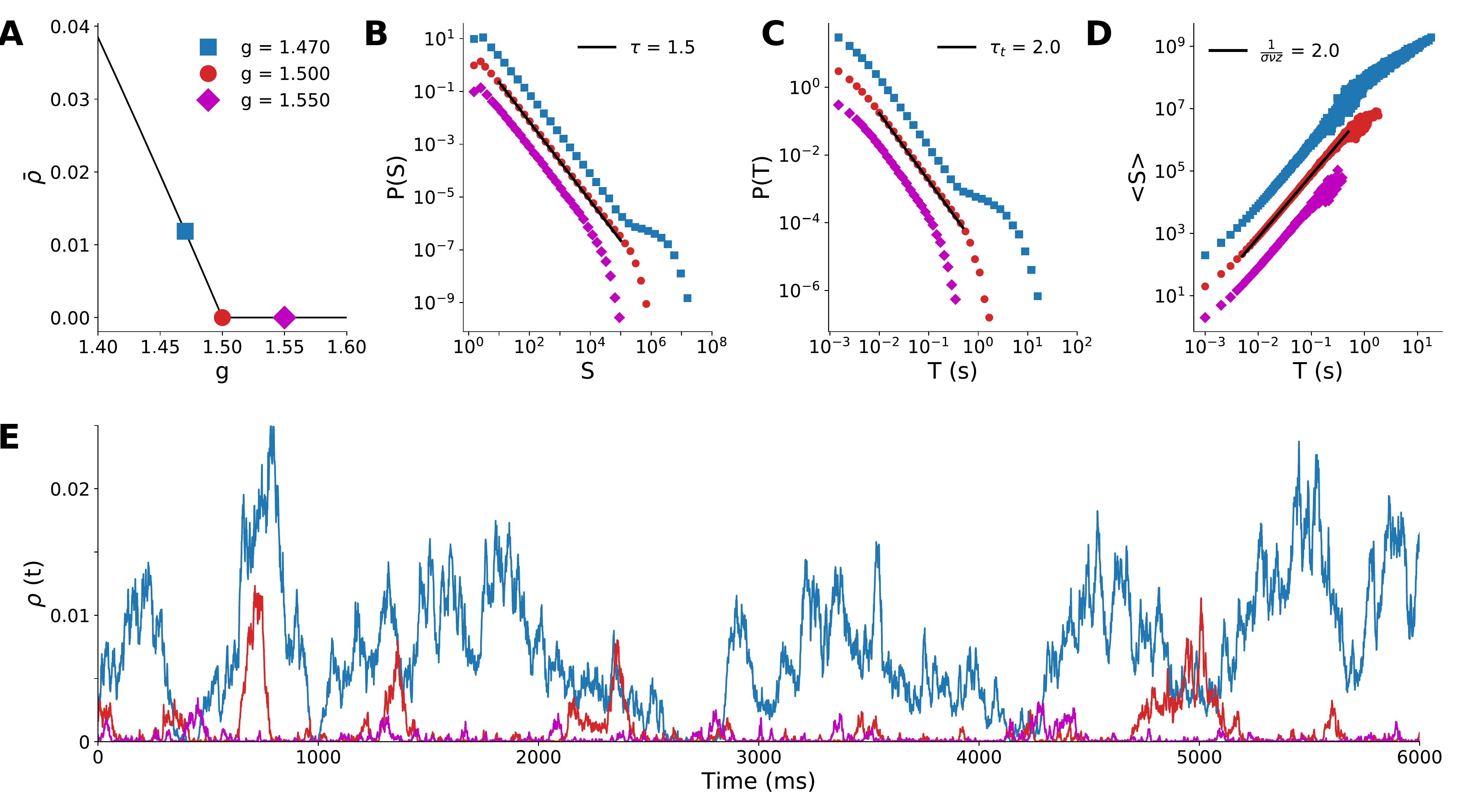}
\caption{Spiking model results with full sampling. 
Behavior of the spiking model ($N=10^5$) for different values of the control parameter $g$. \textbf{(A)} Stationary density of firings $\bar\rho$ as a function of the inhibition
strength $g$ (critical point is the red circle at $g_c=1.5$). 
Solid line is the mean-field solution~\citep{Girardi2020bal}, points are simulation results. 
Distribution of avalanche sizes \textbf{(B)} and duration \textbf{(C)} for the subcritical ($g>g_c$), critical ($g=g_c$) and supercritical ($g<g_c$) regimes.
\textbf{(D)} Average avalanche size $\langle S \rangle$ of a given duration $(T)$. 
\textbf{(E)} Time series of the density of active sites for the three regimes.}
\label{fig:modelo}
\end{figure*}

We start by illustrating the second order phase transition that the model undergoes at a critical value $g_c=1.5$ of the inhibition parameter~\citep{Girardi2020bal}. As shown in Figure~\ref{fig:modelo}A, the stationary density of active sites $\bar\rho$ is  positive for $g<g_c$ (the supercritical regime) and null for $g>g_c$ (the subcritical regime). 

At the critical point $g=g_c$, the distribution of avalanche sizes and duration obey the expected power laws (Equations~\eqref{eq:P(S)} and~\eqref{eq:P(T)}) with exponents $\tau=3/2$ and $\tau_t=2$~\citep{Girardi2020bal}. 
Subcritical avalanches are exponentially distributed, whereas the supercritical distribution has a trend to display large avalanches in space and time (Figure~\ref{fig:modelo}B and C).
Both sides of the scaling law in Equation~\eqref{eq:scaling} independently agree, since the fit to $\langle S \rangle(T)$ yields $1/(\sigma\nu z)=2$ on the critical point (Figure~\ref{fig:modelo}D).
Figure~\ref{fig:modelo}E shows typical time series of firing events for the three regimes. 
These exponents and dynamic behavior of the model is typical of a system
undergoing a MF-DP phase transition.

\subsection{Comparison of subsampled model and experiments stratified by CV}

We now revisit the model by subjecting it to the same constraints that apply to experimental datasets~\citep{fontenele2019criticality} and compare the results between the two.
More specifically: 1) data analysis necessarily uses only a tiny fraction of the total neurons in the system and
2) in urethane-anesthetized rats, cortical spiking variability is a proxy for cortical states~\citep{harris2011cortical} and changes a lot during the hours-long recordings~\citep{clement2008cyclic,Vasconcelos2017}. 

\begin{figure*}[!tbhp]
    \includegraphics[width=\linewidth]{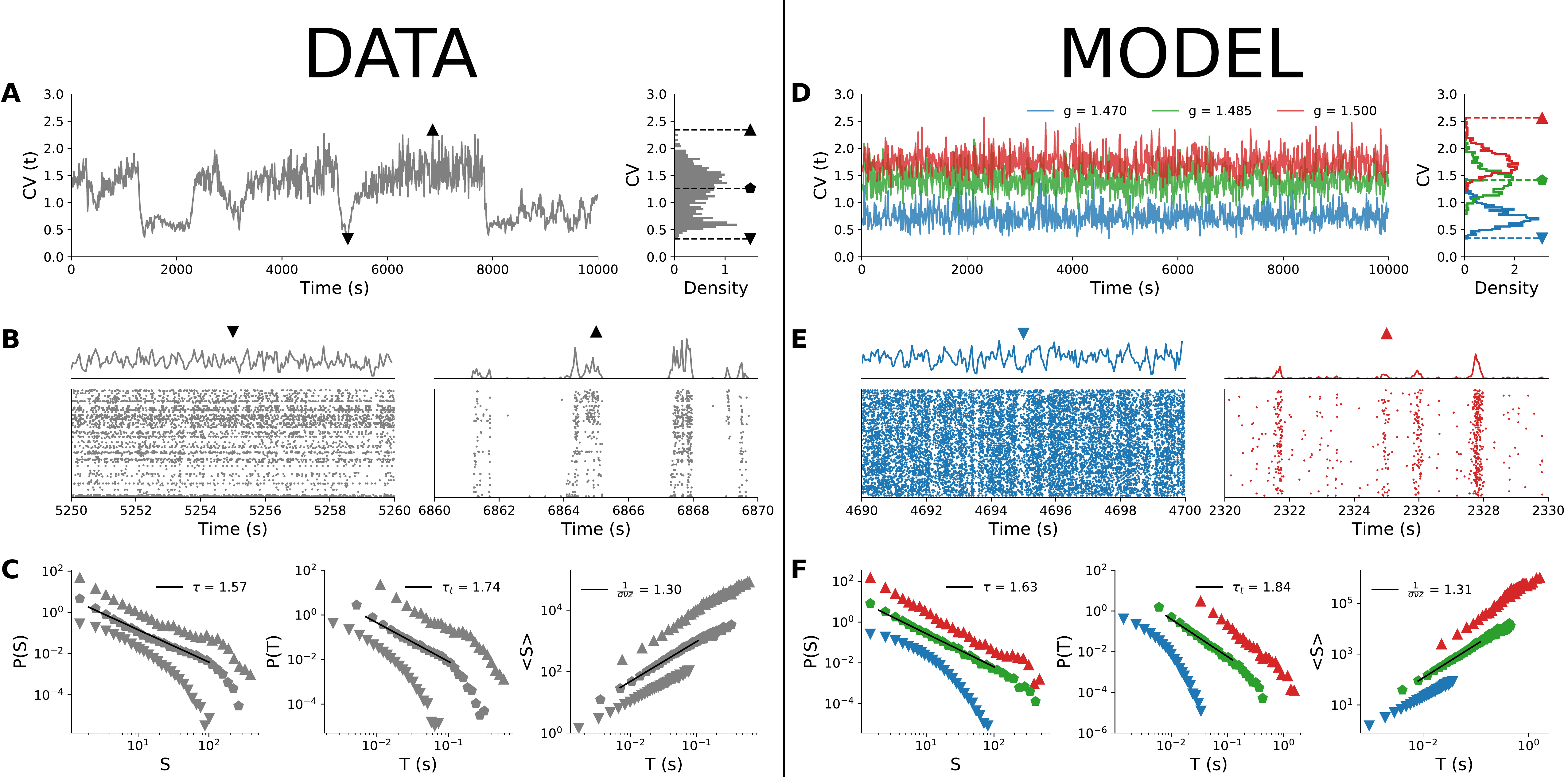}
    \caption{
    Comparison between empirical data and subsampled spiking model.
    $CV$ time series and distribution for \textbf{(A)}
    experimental data (single animal) and \textbf{(D)} model with $n=100$.
    Raster plots and population firing rate in cases of
    low ($\bigtriangledown$) and high ($\bigtriangleup$)
    values of $CV$ for \textbf{(B)} experimental data and \textbf{(E)} model. 
    Scaling exponents $\tau$, $\tau_t$ and $1/(\sigma\nu z)$ for three different values of $CV$ (denoted by different symbols): \textbf{(C)} experimental data and \textbf{(F)} model.
    For both experimental data and model, $w=10$~s.}
    \label{fig:datamodel-raster}
\end{figure*}

Starting with the experimental results, Figure~\ref{fig:datamodel-raster}A shows the time
series of the coefficient of variation ($CV$) of the population spiking activity.
The lowest $CV$ values correspond to asynchronous spiking activity,
whereas the highest values correspond to more synchronized activity 
(both shown in Figure~\ref{fig:datamodel-raster}B).
When we parsed the data by $CV$ percentiles and evaluated neuronal avalanches for different 
percentiles, the distributions varied accordingly,
with exponents $\tau$, $\tau_t$ and $1/(\sigma\nu z)$ varying continuously across
the $CV$ range (Figure~\ref{fig:datamodel-raster}C)
as expected~\citep{fontenele2019criticality}. 

Can the MF-DP spiking network model reproduce these experimental results?
We found that by sampling only a few neurons out of the entire network, indeed it can. 
We sample only $n=100$ neurons out of $N=10^5$, a number that is of the same order as the amount of neurons captured in our empirical data~\citep{fontenele2019criticality}. 
Then, we apply to the subsampled simulation data exactly the same pipeline used for experiments (Section~\ref{sec:cvparsing}).

In the model, we change the E/I level $g$ to control for the variability level $CV$. 
For a fixed value of parameter $g$, $CV$ is a bell-shaped distribution with finite
variance. 
The $CV(t)$ time series of the model for a single $g$ does not present the dynamical complexity observed experimentally (compare Figures~\ref{fig:datamodel-raster}A
and~\ref{fig:datamodel-raster}D).
By varying $g$ within a narrow interval around the critical point $g_c$, the $CV$ distribution of the model covers the values observed experimentally (Figure~\ref{fig:datamodel-raster}D), with less synchronous behavior for low $CV$ and more synchronous activity for high $CV$ (Figure~\ref{fig:datamodel-raster}E; 
the full behavior of the $CV$ distribution as a function of parameter $g$ is shown in Figure~\ref{fig:CVvsg}A). 
Parsing the data by $CV$ and running the avalanche statistics for the subsampled model, we obtain scaling exponents that vary continuously in remarkable similarity to what is observed in the experimental data (Figure~\ref{fig:datamodel-raster}F).


\begin{figure*}[!tbhp]
    \includegraphics[width=\linewidth]{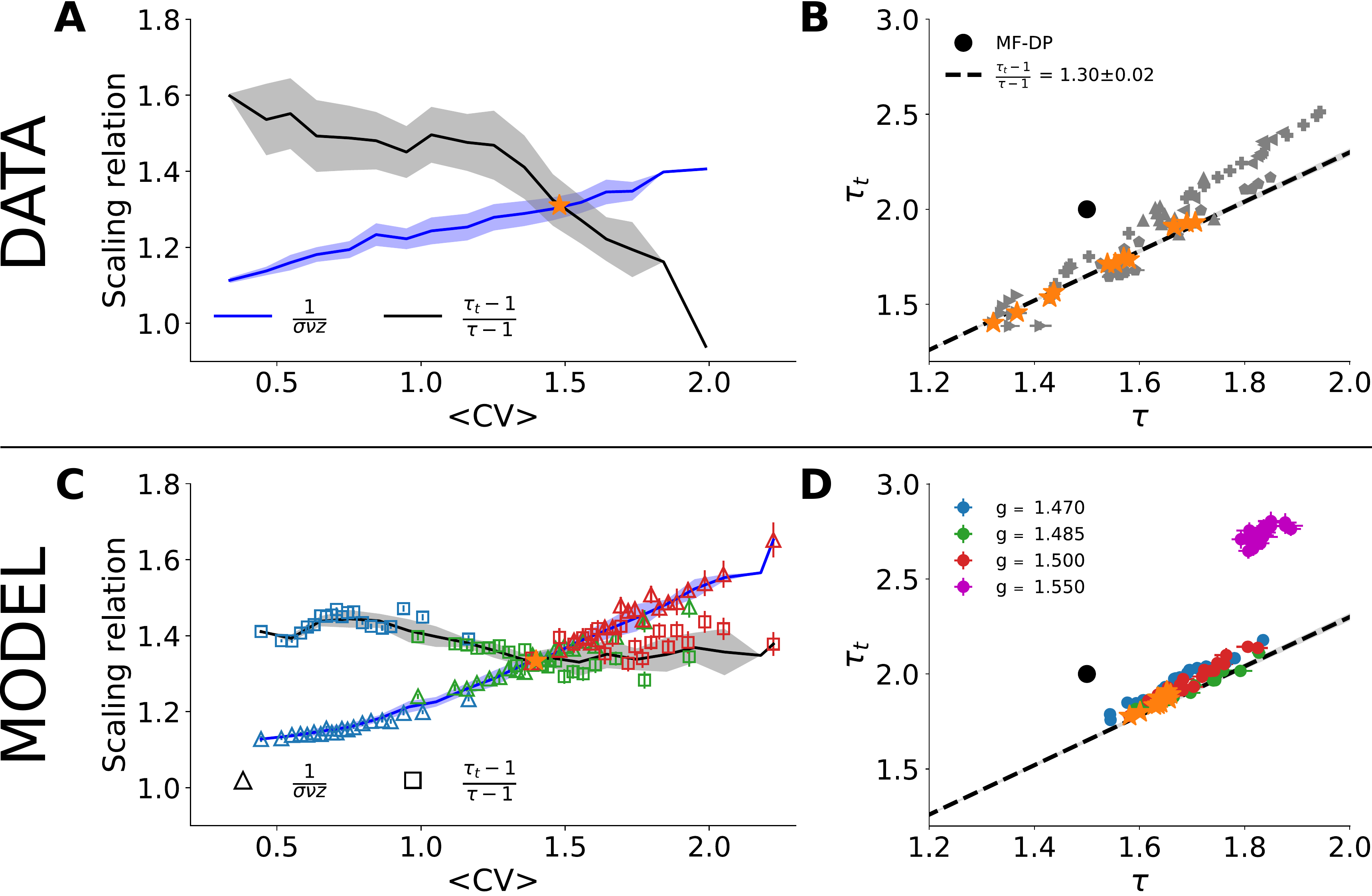}
    \caption{
    Scaling relation and parametric plot of avalanche exponents. 
    Right- and left-sides of Equation~\eqref{eq:scaling} (line and shade are average and standard deviation across the group) as a function of the average $CV$ for \textbf{(A)} experimental data and \textbf{(C)} subsampled model ($n=100$; note that color code and values of $g$ are the same as in Figures~\ref{fig:datamodel-raster}D and~\ref{fig:datamodel-raster}F). 
    Scatter plot in the $(\tau,\tau_t)$ plane for \textbf{(B)} experimental data and \textbf{(D)} subsampled model. 
    In both cases, $\Delta t = \langle ISI \rangle$ and $w=10$~s. 
    The star points in \textbf{(B)} and \textbf{(D)} indicate the values of $\tau$ and $\tau_t$ that satisfied Equation \eqref{eq:scaling} in \textbf{(A)} and \textbf{(C)}.}
    \label{fig:bird}
\end{figure*}


To ensure criticality, we must require that the scaling relation
in Equation~\eqref{eq:scaling} is satisfied.
Figure~\ref{fig:bird}A shows the independent experimental fits
for the left- and right-hand sides of Equation~\eqref{eq:scaling}.
The crossing at $CV_* \simeq 1.46\pm 0.08$ is consistent with a phase transition at an
intermediate point between asynchronous and synchronous
behavior~\citep{fontenele2019criticality}.
In the crossing $CV_*$, we obtain $\tau_*=1.54 \pm 0.12$, $\tau_{t*}=1.73 \pm 0.18$ and
$1/(\sigma\nu z)_*=1.30 \pm 0.02$. 
Plotting $\tau$ versus $\tau_t$, the experimental data scatter along the line with slope
given by $1/(\sigma\nu z)_*$ for different values of $CV$ (Figure~\ref{fig:bird}B).
These results are in agreement with those of Fontenele et al.~\citep{fontenele2019criticality}, again suggesting an incompatibility with the MF-DP universality class.  

The results for the subsampled spiking model, however, suggest otherwise.
We do exactly the same procedure with the subsampled model and find a similar
$CV$ for the crossing of the critical exponents,
${CV_*^{model} \simeq 1.41\pm 0.05}$, when controlling for the E/I ratio $g$ very close to the
critical point $g_c=1.5$ (Figure~\ref{fig:bird}C).
On the crossing $CV_*^{model}$,
we obtain ${\tau_*=1.65 \pm 0.02}$, ${\tau_{t*}=1.87 \pm 0.03}$
and ${1/(\sigma\nu z)_*=1.34 \pm 0.02}$. Note that these critical exponents are
not the real exponents of the model.
In fact, they are apparent exponents generated by subsampling
the network activity. The real critical exponents are
$\tau=3/2$, $\tau_t=2$ and $1/(\sigma\nu z)=2$ (as
we showed in Figure~\ref{fig:datamodel-raster}).

To reproduce the experimental results, the control interval of $g$ was slightly biased
towards the supercritical range: $g_{min} \simeq 1.47 \leq g \leq g_{max} \simeq 1.50$. 
Our model predicts, then, that the whole range of experimental results is produced by
fluctuations of only about 2\% around the critical point
(Figure~\ref{fig:bird}D).
For instance, for $g=1.55$ (3\% above the critical point in the subcritical regime), the scaling relation is no longer satisfied and the measured exponents fall far away from the linear relation observed experimentally in the $(\tau,\tau_t)$ plane (Figure~\ref{fig:bird}D).

This result shows that the MF-DP phase transition under subsampling conditions
is capable of reproducing a whole range of experimentally observed avalanches
due to different $CV$.
To test the robustness of our findings, we employ exactly the same procedure to a simpler model, the probabilistic cellular automaton (Section~\ref{sec:KC}). 
This model is also knowingly of the MF-DP type~\citep{kinouchi2006optimal}, but has a random network topology. 
All the results were similar (see Figure~\ref{fig:AC}), showing that the apparent exponents are a direct consequence of subsampling.

\subsection{Dependence on sampling fraction and time bin width}
\label{sec:dependencesubsampling}

\begin{figure*}[tbhp]
    \includegraphics[width=\linewidth]{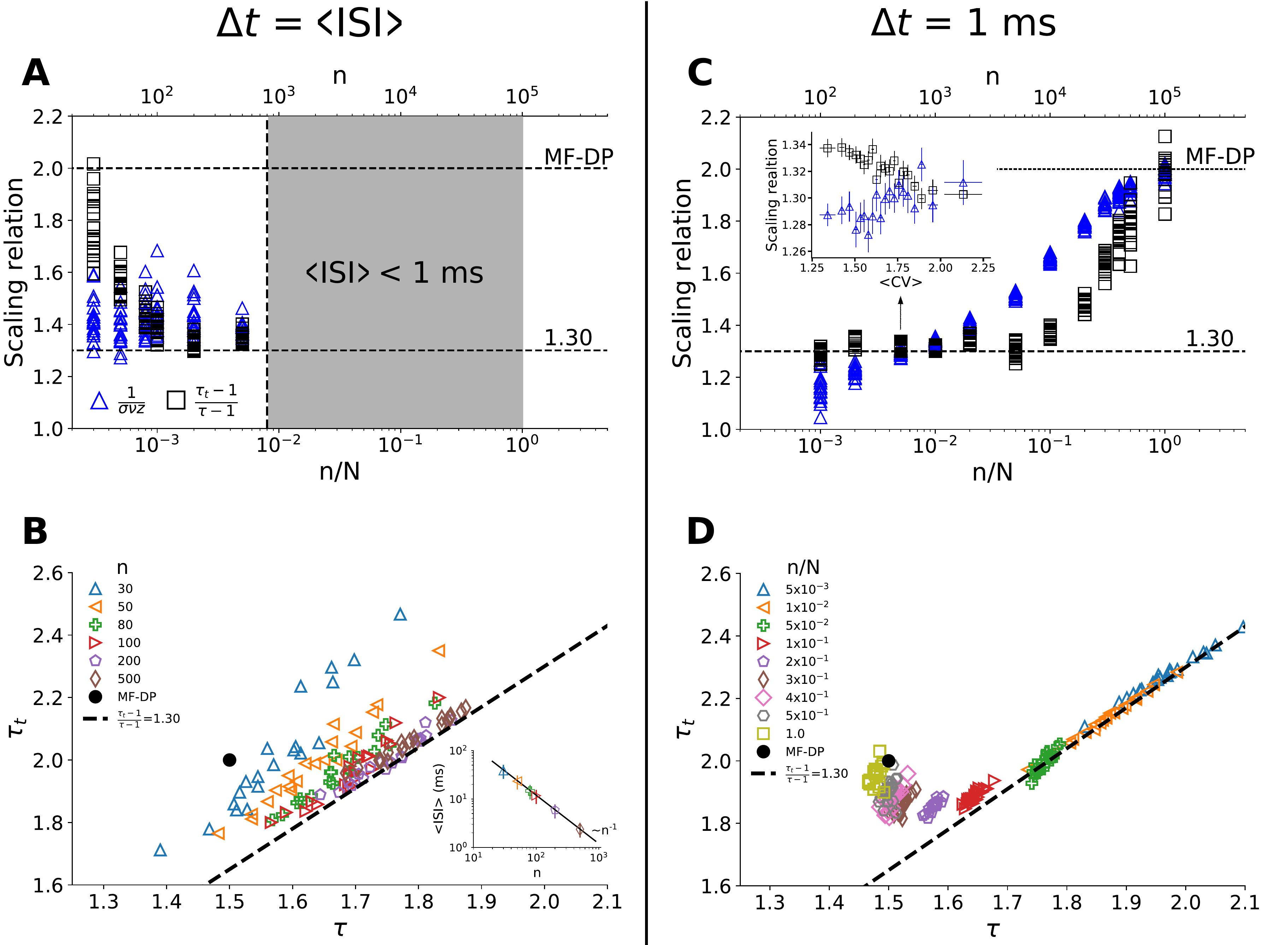} 
    \caption{Dependence of the apparent critical exponents on the sampling parameters.
    In \textbf{(A)} and \textbf{(C)}, we show both sides of the scaling relation (Equation~\ref{eq:scaling}) for \textit{all} values of $CV$ observed in the simulations. For each value of $n/N$, one has the equivalent of the projection of Figure~\ref{fig:bird}C onto its vertical axis. 
    For $\Delta t = \langle ISI \rangle$, \textbf{(A)} the scaling relation is satisfied for increasing number of sampled neurons \textbf{(B)} with exponents that agree with experimental data. 
    Since $\langle ISI \rangle$ decreases with $n$ (inset of \textbf{(B)}), this analysis breaks down when $n$ is so large that $\langle ISI \rangle$ becomes smaller than 1~ms  (grey region in \textbf{(A)}), which is the time step of the simulations. 
    For $\Delta t = 1$~ms, \textbf{(C)} the scaling relation is satisfied for small $n/N$, within a relatively wide range of $CV$ values (inset of \textbf{(C)}). 
    For $n/N\to 1$, results converge to MF-DP values (\textbf{(C)} and \textbf{(D)}), as expected. Simulations with $N=10^5$ and $g=1.5$.
}
    \label{fig:subsampling}
\end{figure*}

How robust are the results of the model against variation in the sampling size ($n$) and time bin width ($\Delta t$)?
First, we consider the time bin width as
the population interspike interval $\Delta t=\langle ISI\rangle$.
The minimum sampling size we employ is $n=30$ so that power laws still
satisfy Akaike's Information Criterion.
The agreement of both sides of the scaling law enhances with growing sampling fraction (Figure~\ref{fig:subsampling}A and B).
However, $\langle ISI\rangle$ decreases with the number of neurons sampled (inset of Figure~\ref{fig:subsampling}B).
When the natural bin decreases below 1~ms (the time step of the model), the analysis no longer makes sense.
As $n$ increases, the relation between $\tau$ and $\tau_t$ converges to the apparent critical scaling that fits experimental results~(Figure~\ref{fig:subsampling}B). 


To check whether we recover the MF-DP real exponents from their apparent values as $n$ increases, we choose the smallest time bin possible, $\Delta t=1$~ms.
We observe that for a small fraction of sampled units ($n/N \sim \mathcal{O}(10^{-2})$)
the scaling relation (Equation~\ref{eq:scaling}) is 
satisfied~(Figure~\ref{fig:subsampling}C) with the apparent critical exponents
that match the experimental results~(Figure~\ref{fig:subsampling}D). 
In fact, the scaling relation in Equation~\ref{eq:scaling} is satisfied for a range of
$CV$ values (inset of Figure~\ref{fig:subsampling}C).
Increasing the sampling further ($n/N \sim \mathcal{O}(10^{-1})$),
the scaling relation ceases to be satisfied (Figure~\ref{fig:subsampling}C) and the avalanche
exponents get separated from the experimental scaling relation
(Figure~\ref{fig:subsampling}D). 
But as $n\to N$, the MF-DP scaling relation is recovered (as it should). 

We have further tested the robustness of these findings by varying the time bin width
used to defined avalanches ($0.75 \leq \Delta t/\langle ISI \rangle \leq 2$).
We observed that experiments and model have very similar behavior
(Figures~\ref{fig:robustness}A and~\ref{fig:robustness}B).
Furthermore, both model and experiments are virtually insensitive
to the width of the $CV$ window $w$
(Figures~\ref{fig:robustness}C and~\ref{fig:robustness}D).

\subsection{Pairwise correlation structure}
\label{sec:correlation}

\begin{figure*}[!tbhp]
    \includegraphics[width=\linewidth]{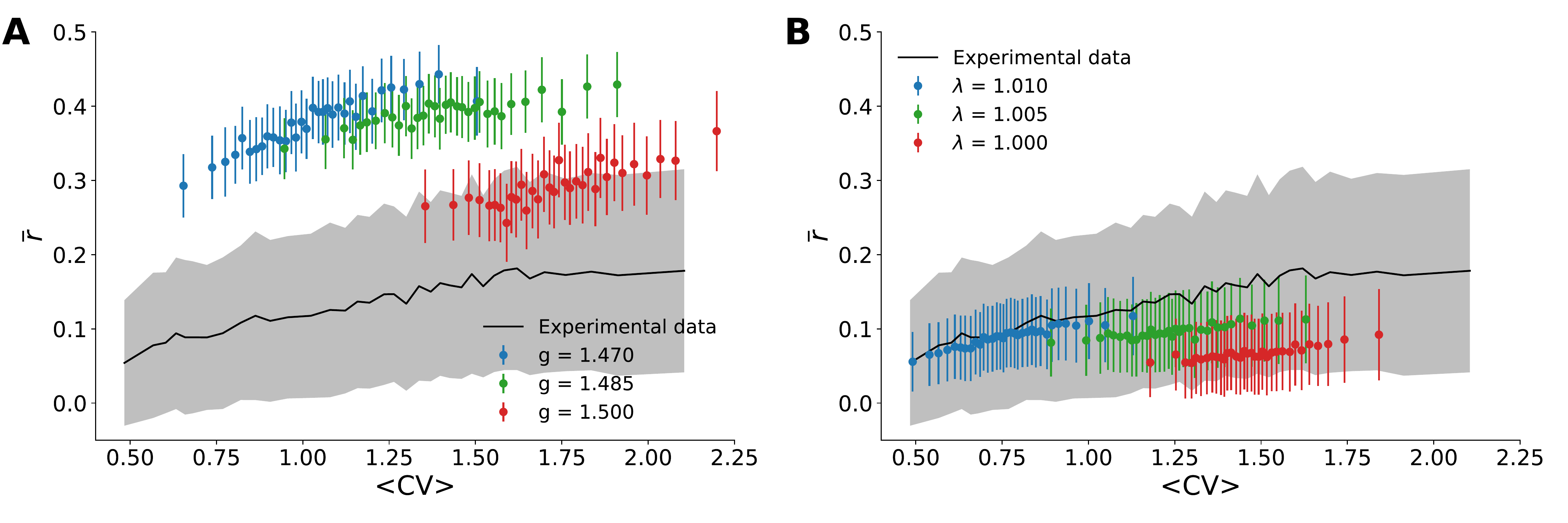} 
    \caption{Correlation structure. The experimental pairwise correlation of firing rates is shown as a function of $\langle CV\rangle$ (black line is the average $\bar{r}$, while gray shading is the standard deviation of the distribution). It is compared with 
    theoretical results for \textbf{(A)} the spiking neuronal network with $n=100$ sampled neurons, and \textbf{(B)} the cellular automaton model with $n=500$ sampled sites. 
    }
    \label{fig:correlationstructure}
\end{figure*}

We also tested the correlation structure of the model and compared it to experimental results.
In the literature on cortical states, asynchronous states are associated with pairwise spiking correlations $r^{(k,l)}$ which are distributed around an average $\bar{r}$ close to zero, whereas synchronous states have positive average~\citep{harris2011cortical}. 
This is quantified in Figure~\ref{fig:correlationstructure}A, where $\bar{r}$ is shown to increase monotonically with $CV$.
For the experimental data, $\bar{r}$ reaches zero within the standard deviation of the distribution for sufficiently small $CV$. 

Compared with the experimental results, the spiking model with inhibition generally overestimates $\bar{r}$ (Figure~\ref{fig:correlationstructure}A). 
This could be due to its all-to-all connectivity. 
The cellular automaton model on a random graph yields quantitatively better results  (Figure~\ref{fig:correlationstructure}B). 
In either case, we observe again that, just like for the scaling relation (Figure~\ref{fig:bird}), the correlation structure of the experimental data is relatively well reproduced by very small deviations around critical parameter values. 

\section{Discussion and conclusions}

We revisited the results recently published by Fontenele et al.~\citep{fontenele2019criticality} by repeating their analyses on new experimental data and two different models.
To test the idea that the urethanized cortex hovers around a critical point,
we stratified the avalanche analyses across cortical states.
For the new experimental data, we verified that the scaling relation combining the three
exponents (Equation~\ref{eq:scaling}) is indeed satisfied at an intermediate value $CV_*$,
away from the synchronous and asynchronous extremes.
At this critical value, the three exponents differ from those of the MF-DP universality class,
thus confirming previous findings~\citep{fontenele2019criticality}. 

We addressed whether the exponents of the MF-DP universality class and those observed
experimentally could be reconciled, despite their disagreement.
In other words, we return to the question: if the brain is critical, what is the phase
transition? Do the experimental results presented here and in Fontenele et al.~\citep{fontenele2019criticality}
refute branching-process-like models as explanations? 


To answer these questions, we relied on two models: an E/I spiking neuronal network in an all-to-all graph; and a probabilistic excitable cellular automaton in a random graph. 
Despite the simplicity and limitations of these models (which we discuss below), they have a fundamental strength that led us to choose them:  they are very well understood analytically.  
In both cases, mean-field calculations agree extremely well with simulations, so that we are safe in locating the critical points of these models~\citep{Girardi2020bal,kinouchi2006optimal}. 
This is very important for our purposes, because it allows us to test whether the models can reproduce the data and, if so, how close to the critical point they have to be. 
Besides, their universality class is also well determined:
the exponents shown in Figures~\ref{fig:modelo}B,~\ref{fig:modelo}C~and~\ref{fig:modelo}D are those of with MF-DP. 


The crucial point is that the results in Figure~\ref{fig:modelo} are based on avalanches which are measured by taking into account \textit{all} simulated units of the model, a methodological privilege that is not available to an experimentalist measuring spiking activity of a real brain with current technologies. 
In fact, a considerable amount of work has shown that subsampling can have a drastic effect on the avalanche statistics of models~\citep{Priesemann09,ribeiro2010spike,Girardi2013, Ribeiro14a, Priesemann2014subcriticalbrain,Levina2017subsampling,Wilting2019criticality}. 
Therefore, here we set out to test whether MF-DP models could yield results nominally incompatible with that universality class if they were analyzed under the same conditions as the data, i.e. with $CV$ parsing and severe subsampling. 


Both subsampled models can quantitatively and qualitatively reproduce the central features
of the experimental results. The scaling relation (Equation~\ref{eq:scaling}) is satisfied
at an intermediate value $\langle CV\rangle_*$, with the correct qualitative behavior of both
sides of the equation: $1/(\sigma\nu z)$ increases with $CV$, while $(\tau_t-1)/(\tau-1)$ decreases (see Figures~\ref{fig:bird}A,~\ref{fig:bird}C and~\ref{fig:AC}D).
In fact, the values of $\langle CV\rangle_*$, and those of the apparent exponents of the
MF-DP models, $\tau_*$, $\tau_{t*}$, and $1/(\sigma\nu z)_*$, agree with
the experiments within errorbars.
Moreover, even away from the point $\langle CV\rangle_*$ where Equation~\eqref{eq:scaling} is satisfied, the spread of the exponents $\tau$ and $\tau_t$ of the subsampled models follows an almost linear relation (Figures~\ref{fig:bird}D and~\ref{fig:AC}E), in good agreement with not only our experimental results (Figure~\ref{fig:bird}B), but also with those of other experimental setups~\citep{fontenele2019criticality}. 
When we sample from the whole network, we recover the true critical exponents of the model, confirming that spatial subsampling and temporal binning are sufficient ingredients to push the urethanized rats' brains critical exponents toward apparent values, hiding its putative true critical phase transition.



Knowing analytically the critical points of the models, we can check in which parameter range they successfully reproduce the experimental results. 
As it turns out, the scaling relation and the linear of spread of exponents are reproduced by the subsampled models only if they are tuned within a narrow interval around their critical points. 
The model still fits well the urethanized cortex data up to 3\% away towards the supercritical state.
This corroborates the hypothesis that the brain operates in a quasicritical state~\citep{Bonachela2010} that is slightly supercritical~\citep{Costa2017Entropy,Girardi2020bal}. 
This contrasts with claims that the brain might operate slightly subcritically~\citep{Priesemann2014subcriticalbrain}.
Note that if the model becomes too subcritical, the size and duration exponents fall very far apart from the experimentally observed linear relation (Figure~\ref{fig:bird}D). 
If it is too supercritical, there are not enough silent windows to distinguish avalanches in the first place.
This discrepancy is a topic for further investigation, since our models have operated in the limit of zero external field (i.e., external stimuli was only employed to generate an avalanche). Priesemann et al.~\citep{Priesemann2014subcriticalbrain}, for instance, tried to follow a balance between internal coupling and external stimulus
in order to maintain an average firing rate.
Whether that different approach would lead to significant changes to our 
results remains to be investigated. 

Despite the small variation of the model E/I levels controlled by $g$, the variation of $CV$ is large enough to essentially cover the range of experimentally observed values (Figures~\ref{fig:datamodel-raster}A,~\ref{fig:datamodel-raster}D~and~\ref{fig:AC}B). 
This is due in part to the fact that we evaluated $CV$ within finite windows of
width $w=10$~s. 
In Figure~\ref{fig:sigmaCV1}, we show that the standard deviation of $CV$ is a
decreasing function of the time used to estimate it.
This implicates that a better resolution for
experimental $CV$ can be obtained by increasing this time windows to 20~s. 
It is important to note, however, that in experiments one needs to reach
a good trade-off
between a better statistical definition of $CV$ and not mixing different 
cortical states due to the nonstationarity characteristic of the urethane 
preparation (as depicted in Figure~\ref{fig:datamodel-raster}A).

Perhaps even more important than the range of $CV$ values obtained around the critical point of the models is the richness of the experimentally observed temporal evolution of $CV$ (Figure~\ref{fig:datamodel-raster}A).
The model needs to be fine tuned to different values of E/I levels in order to get different average values of $CV$.
This is one of the limitations of the models which would be worth
addressing next.
One possibility would be to replace static models (i.e., with fixed control 
parameters) with ones with plasticity, in which coupling parameters are 
themselves dynamic variables and the critical point is obtained via 
quasicritical self-organization~\citep{Costa15,Brochini2016,Campos2017correlations,Costa2017Entropy,Kinouchi2019stochastic,Girardi2020bal}. 

Another limitation of the models is their simple topology, which in future 
works could be improved to come closer to cortical 
circuitry~\citep{PotjansDiesmann}. 
This would likely come at the cost of foregoing analytical results to start 
with, thus augmenting the computational efforts involved. But it would 
certainly allow to probe the robustness of the results presented here against 
more realistic topologies.
On the other hand, there is quantitative agreement
between the apparent exponents of both models
(each having a different topology) with the experimental
exponents. This suggests that at the scale of the present phenomenology,
the average topology should play a minor role.


The fact that subsampling seems to be a crucial ingredient for explaining the 
data is a double-edged sword. 
On the one hand, it allowed us here to reconcile MF-DP models with results for
spiking data in the anesthetized rat cortex. 
On the other hand, note that even measurements which should in principle be less
prone to subsampling, such as LFP results in the visual cortex of the turtle~\citep{Shew15}, still fall on the same scaling line of $\tau$ versus $\tau_t$ (Figure~\ref{fig:bird}B) as those of spiking
data~\citep{fontenele2019criticality},
both having apparent non-MF-DP critical exponents.
~This issue is not addressed by the current model and deserves further
investigation. 
Our results point only to MF-DP models as sufficient,
not as necessary, to explain the observed phenomenology.
So it is at least conceivable that different models with different phase
transitions~\citep{diSanto2018LandauGinzburg,DallaPorta2019,Pinto2019oscillations}
could also yield non-trivial true or apparent exponents compatible with the
data, even without subsampling~\citep{fontenele2019criticality}.

Finally, our simulation results underscore the methodological vulnerabilities
of assessing criticality exclusively via avalanche analysis. 
Not only are MLE power-law fits sensitive to parameters but even a more
stringent scaling analysis can lead to non-trivial
apparent exponents which are an artifact of subsampling,
as we have shown. 
Therefore, the development of additional figures of merit, such as control and
order parameters, susceptibilities and
others~\citep{Yang2012maximal,Tagliazucchi12,Yu2013universal,tkavcik2015thermodynamics,mora2015dynamical,Girardi2016,GirardiV12018,lotfi2020signatures},
remains a very important line of research to strengthen studies of brain criticality. 

\bigskip 

\section{Acknowledgment}

The authors acknowledge Fundação de Amparo \`a Ciência e Tecnologia de Pernambuco (FACEPE) (grants APQ-0642-1.05/18 and APQ-0826-1.05/15), 
Universidade Federal de Pernambuco, 
Conselho Nacional de Desenvolvimento Científico e Tecnológico (CNPq, grants 425329/2018-6 and 301744/2018-1),
Coordenação de Aperfeiçoamento de Pessoal de Nível Superior (CAPES) and FAPESP (grant 2018/09150-9).
This paper was produced as part of the activities of Research, Innovation and Dissemination Center for Neuromathematics
(grant No. 2013/07699-0, S. Paulo Research Foundation -- FAPESP).
We thank Osame Kinouchi for discussions and a critical read of this manuscript.

\bibliographystyle{ieeetr}

\appendix

\section{CV distribution as a function of model parameters}
\label{sec:CVdistribution}
\renewcommand{\thefigure}{A\arabic{figure}}

\setcounter{figure}{0}

\begin{figure}[tbhp]
    \begin{center}
    \includegraphics[width=\linewidth]{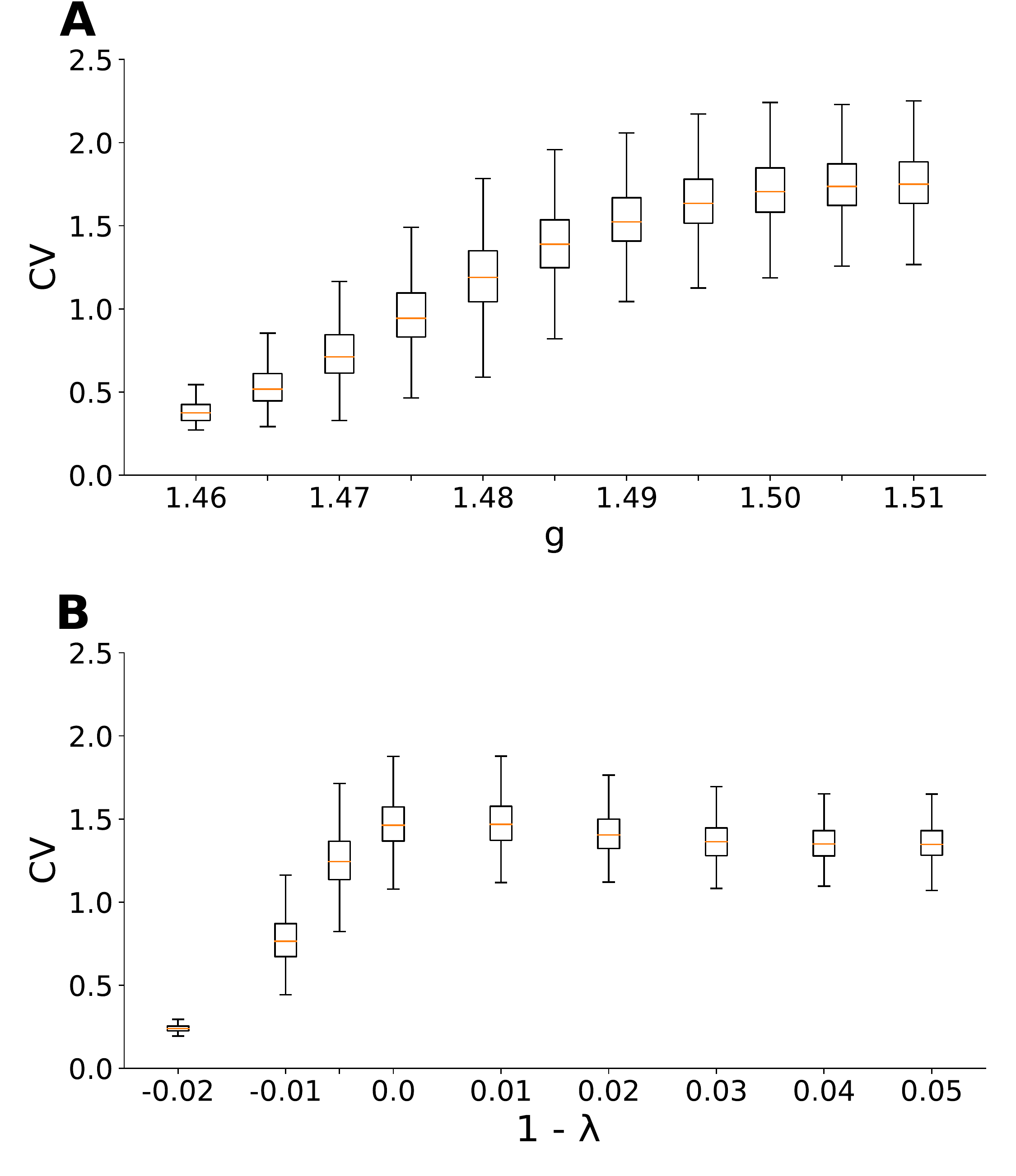} 
    \end{center}
     \caption{
     $CV$ distribution versus model parameters. 
     Boxplots of $CV$ as a function of  parameter \textbf{(A)} $g$ for the spiking model with excitation and inhibition and \textbf{(B)} $1-\lambda$ for the probabilistic cellular automaton model.  
     In both cases, $w=10$~s, just like for experimental data. 
     \textbf{(A)} $n=100$. \textbf{(B)} $n=500$.
     }
     \label{fig:CVvsg}
\end{figure}

Here we illustrate the distribution of $CV$ values as parameter values of the models are varied. 
In both cases, $CV$ was estimated like in experimental data, i.e. for a subsampled number of units and during a finite window of time. 
Note that, as the net excitation decreases along the horizontal axes in Figure~\ref{fig:CVvsg}, the average values of $CV$ initially increase, then saturate.

\section{Results for the cellular automaton model}
\label{sec:appKC}
\renewcommand{\thefigure}{B\arabic{figure}}

\setcounter{figure}{0}

\begin{figure*}[tbhp]
    \includegraphics[width=\linewidth]{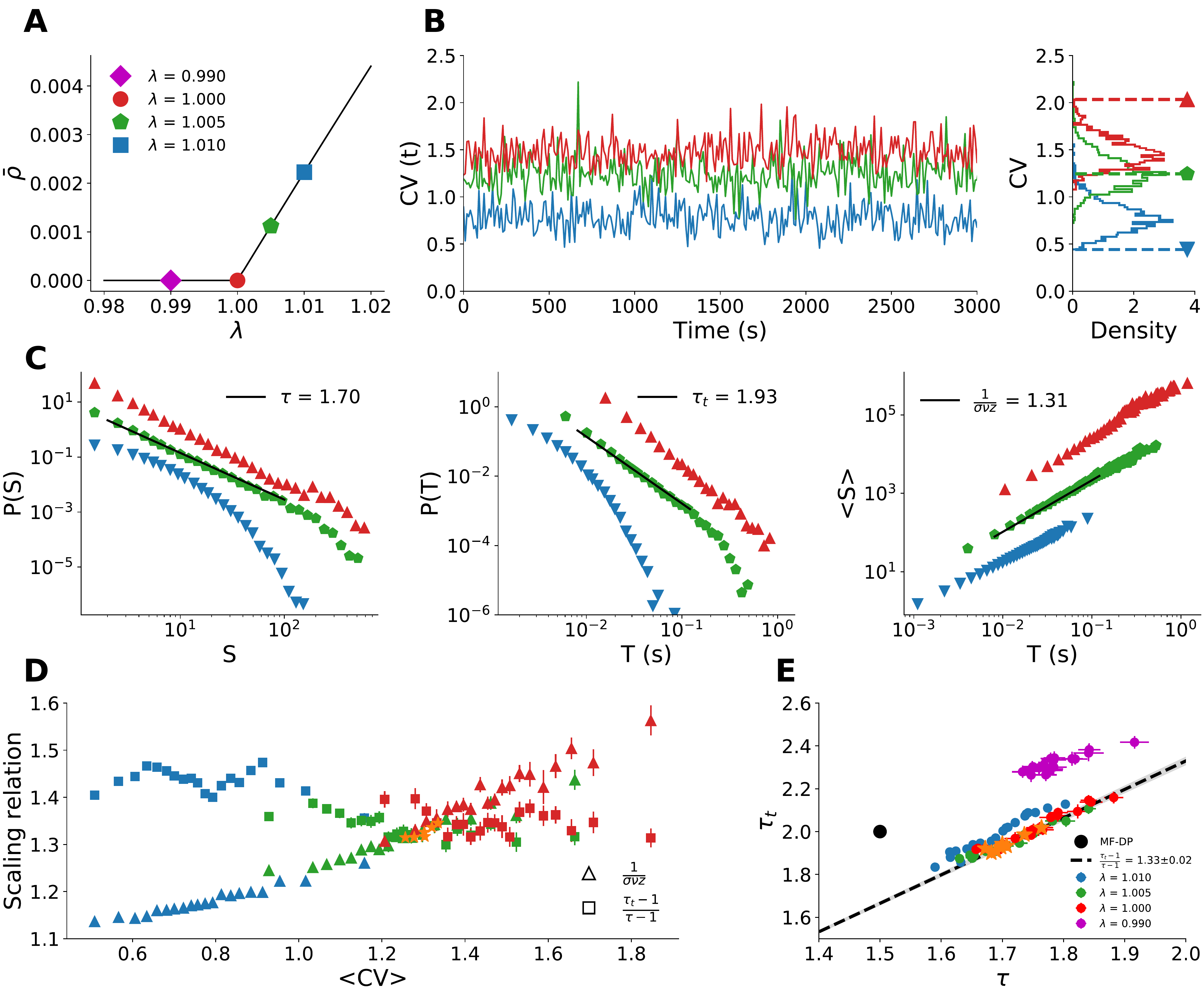} 
    \caption{A probabilistic cellular automaton model with excitation only.
    \textbf{(A)} Stationary density of active sites as a function of the control parameter (branching ratio) $\lambda$ for the fully sampled model with $N=10^5$ (see Section~\ref{sec:KC}). Points are simulations and lines are the linear expansion of the mean-field solution.
    All the remaining plots are for the subsampled model with
    $n = 500$. 
    \textbf{(B)} $CV$ time series and histogram around the critical point $\lambda_c=1$. 
    \textbf{(C)} Exponents $\tau$, $\tau_t$ and $1/(\sigma\nu z)$ depend on $\lambda$. 
    \textbf{(D)} Right- and left-hand sides of the scaling relation Equation~\eqref{eq:scaling} coincide around $CV_*^{CA-model} = 1.30 \pm 0.05$. 
    (e) Spread of exponents $\tau$ and $\tau_t$ around the slope $1/(\sigma\nu z)_*^{CA-model}=1.33\pm0.02$. 
    }
    \label{fig:AC}
\end{figure*}

Simulations for the cellular automaton model described in Section~\ref{sec:KC} of the main paper yielded results similar to those obtained for the spiking model. 
Figure~\ref{fig:AC} shows the same plots as in Figures~\ref{fig:datamodel-raster}~and~\ref{fig:bird}, with similar values of the exponents and of the spiking variability at which the scaling relation in Equation~\eqref{eq:scaling} is satisfied: at ${CV_*^{CA-model} = 1.30 \pm 0.05}$, 
${\tau_*^{CA-model}=1.71 \pm 0.03}$,  ${\tau_{t*}^{CA-model}=1.94 \pm 0.03}$ and ${1/(\sigma\nu z)_*^{CA-model}=1.33 \pm 0.02}$.

Note also the same tendency of the model to reproduce the data for slightly supercritical values with the coupling parameter ranging from $1.00 \leq \lambda \leq 1.01$. 
Fluctuations around the critical point $\lambda_c=1$ in parameter space are therefore in the range of 1\%. 

\section{Robustness with respect to time bin}
\renewcommand{\thefigure}{C\arabic{figure}}

\setcounter{figure}{0}

In Figure~\ref{fig:subsampling} of the main text, we explore how the exponents $\tau$ and $\tau_t$ depend on the number of sampled units $n$ and  the choice of two different bins for the analysis of avalanches: $\Delta t = \langle ISI \rangle$ and $\Delta t = 1$~ms. 
Here, we probe further the robustness of the results for experimental and subsampled model data in the evaluation of $\langle CV \rangle _*$, $\tau_*$, and $\tau_{t*}$ that satisfy the criticality criterion (Equation~\ref{eq:scaling}) by assesssing their dependence on $\Delta t$ (in multiples of $\langle ISI \rangle$) and the time window $w$ used to evaluate $CV$.

\begin{figure*}[tbhp]
    \includegraphics[width=\linewidth]{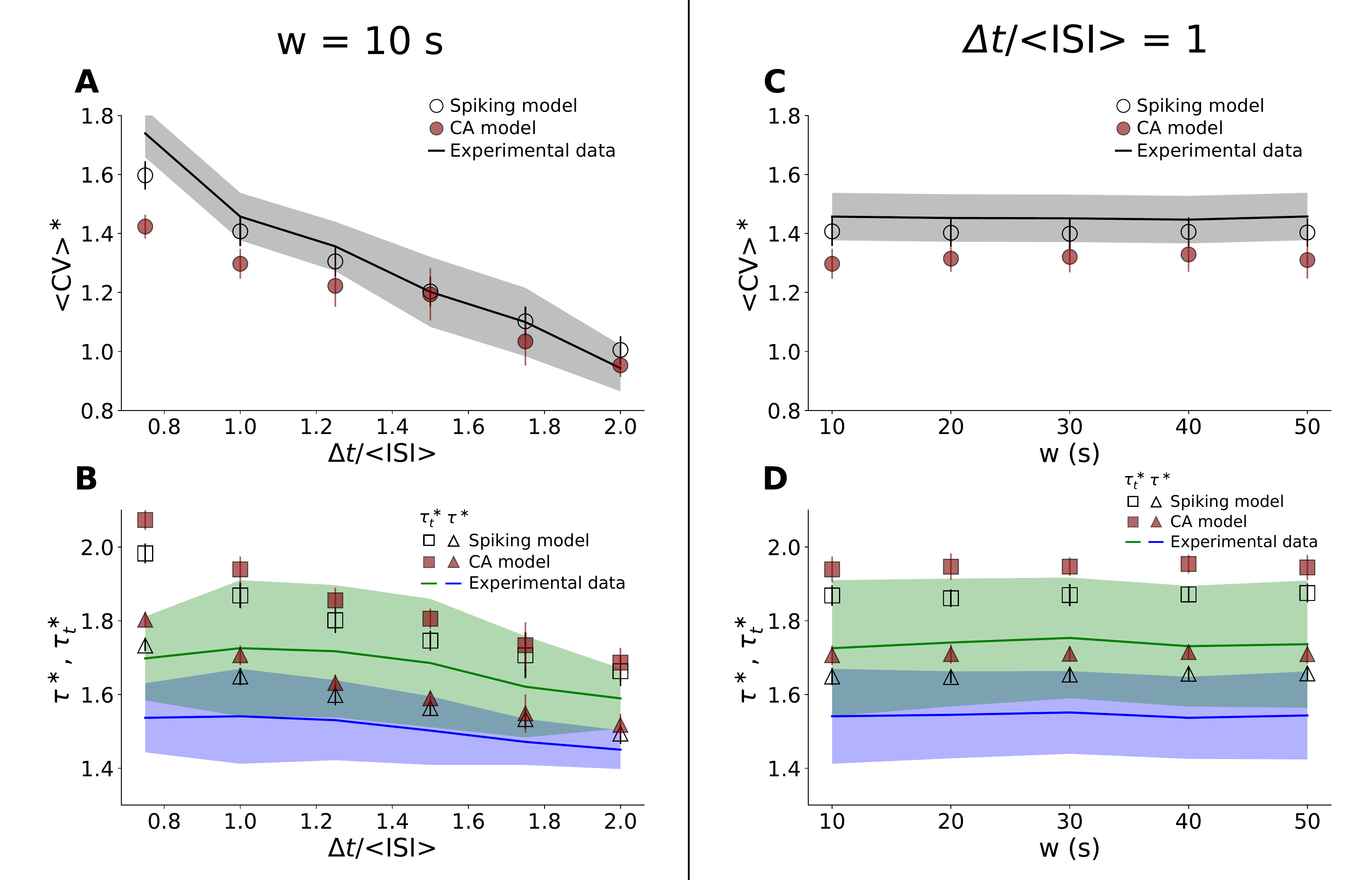} 
    \caption{
    Dependence of $\langle CV \rangle _*$, $\tau_*$ and $\tau_{t*}$ on the temporal windows $w$ and $\Delta t$. 
    We compared the group analysis for the experimental data with results for the subsampled models ($n = 100$ for the spiking model and $n = 500$ for the cellular automaton model). 
    For $w = 10$~s, we evaluated \textbf{(A)} $\langle CV \rangle _*$ and \textbf{(B)} $\tau_*$ and $\tau_{t*}$ at the point where Equation~\eqref{eq:scaling} is satisfied, varying the time bin $\Delta t$ used to calculate avalanches. 
    Next we fixed $\Delta t = \langle ISI \rangle$ and varied $w$, the time window to calculate $CV$, and we evaluated \textbf{(C)} $\langle CV \rangle _*$ and \textbf{(D)} $\tau_*$ and $\tau_{t*}$ at the point where Equation~\eqref{eq:scaling} is satisfied. 
    We noticed that in all these scenarios, the results from the subsampled models follow the behavior of the experimental data. 
    For the spiking model, $g$ was varied from 1.47 to 1.50 and for the cellular automaton model, $\lambda$ was varied from 1.00 to 1.01 (for both models, $N = 10^5$). In \textbf{(C)} and \textbf{(D)}, we use $NB = \{50,25,16,12,10\}$ for $w = \{10,20,30,40,50\}$~s to keep the total sampling time approximately the same (see Methods section in the main paper).}
    \label{fig:robustness}
\end{figure*}

As shown in Figures~\ref{fig:robustness}A and~\ref{fig:robustness}B, both for the experimental data and for the subsampled models, $\langle CV \rangle _*$, $\tau_*$, and $\tau_{t*}$ decrease with the increase of the time bin $\Delta t$, a result which is consistent with those originally obtained by Beggs and Plenz~\citep{beggs2003neuronal}. 
In Figures~\ref{fig:robustness}C and~\ref{fig:robustness}D, $\langle CV \rangle _*$, $\tau_*$, and $\tau_{t*}$ do not suffer great deviations, being largely insensitive to $w$.

\section{CV as a proxy of cortical state}
\renewcommand{\thefigure}{D\arabic{figure}}

\setcounter{figure}{0}

\begin{figure}[tbhp]
    \begin{center}
    \includegraphics[width=\linewidth]{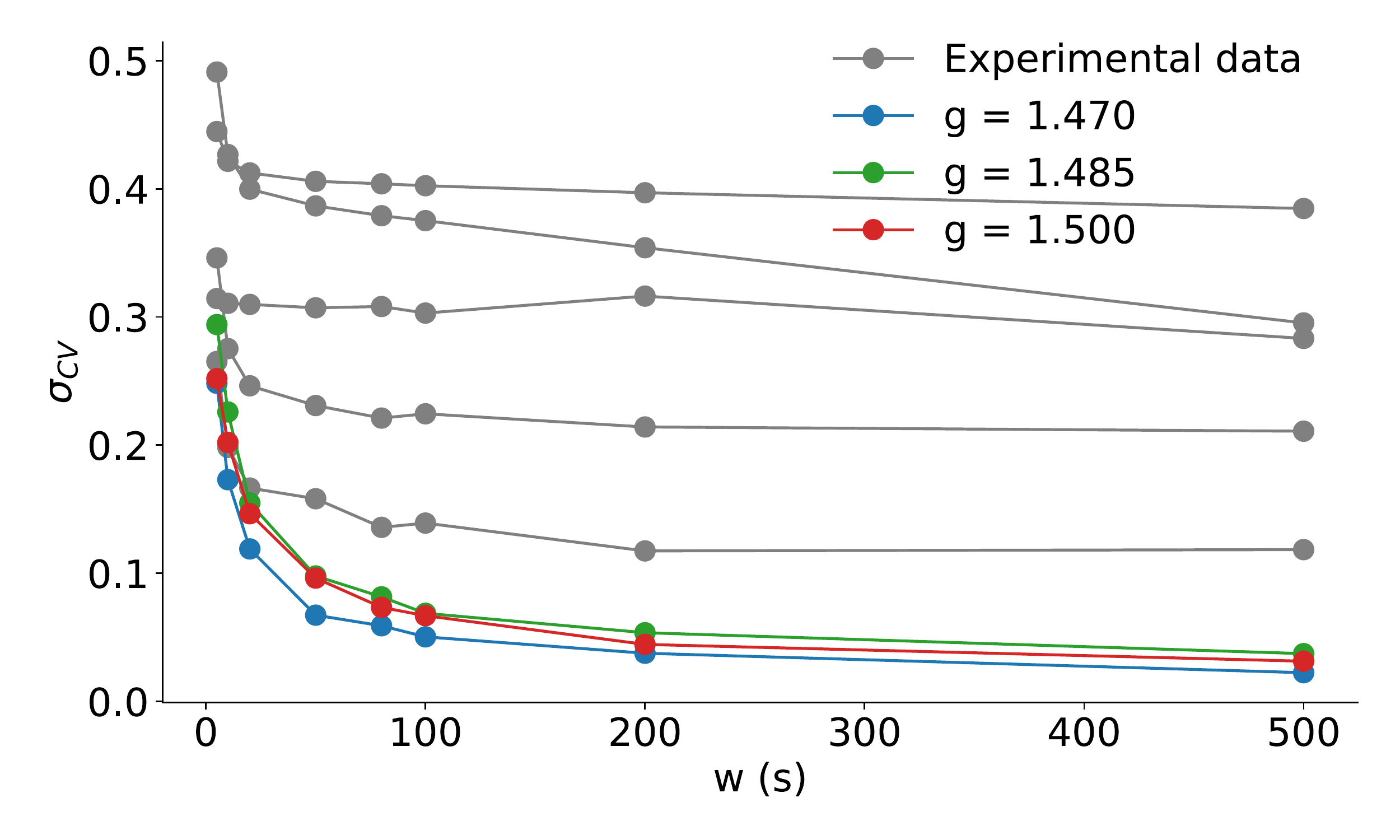} 
    \end{center}
    \caption{Dependency of the standard deviation ($\sigma_{CV}$) of the $CV$ time series with the time window $w$ used to estimate it. Each gray curve represents the result for each rat studied. The colored curves (see legend) represent values of $g$ of the subsampled spiking model with $n = 100$.}
    \label{fig:sigmaCV1}
\end{figure}

It is known in the neuroscience literature that different levels of spiking variability are related to different cognitive states. 
In urethane anesthetized brains, there is a slow modulation of the level of synchronization of the ongoing activity.
From the experimental perspective, since we cannot ensure stationarity, it is preferable to use the minimum necessary time to calculate $CV$ and define a cortical state. 
In the literature it is typically arbitrarily accepted to use $w=10$~s for the estimation of a cortical state.
Here we can evaluate, from the model perspective, if this value of $w$ can already provide a good estimation of $CV$. 
In Figures~\ref{fig:sigmaCV1}, we evaluate the standard deviation $\sigma_{CV}$ of $CV$ as a function of the time used to estimate it. 
As we can see, a time bin of 20~s provides a better discrimination between the cortical states, saturating the decay the experimental CVs standard deviation.
However, despite the fact that a change in $w$ does not impact the results (Figures~\ref{fig:robustness}C~and~\ref{fig:robustness}D), in experiments one needs to compromise between a better statistical definition of $CV$ and not mixing different states due to nonstationarity.

\end{document}